\documentclass[epj]{svjour}
\usepackage{amsfonts}
\usepackage{amsmath}
\usepackage{graphicx}
\usepackage{subfigure}
\usepackage{dcolumn}
\usepackage{bm}
\usepackage{booktabs}
\usepackage{color}

\newcommand{\figcaption}{\def\@captype{figure}\caption}
\newcommand{\tabcaption}{\def\@captype{table}\caption}

\newcommand{\Rmnum}[1]{\expandafter\@slowromancap\romannumeral #1@}
\def\hlinewd#1{%
  \noalign{\ifnum0=`}\fi\hrule \@height #1 \futurelet
   \reserved@a\@xhline}
\makeatother

\def\FF(s){\left[(\alpha+\beta)m_c^2-\alpha\beta s\right]}
\def\HH(s){\left[m_c^2-\alpha(1-\alpha) s\right]}

\allowdisplaybreaks[4]

\usepackage{ulem}

\usepackage{graphics}

\begin{document}

\title{Searching for hidden-charm baryonium signals in QCD sum rules}

\author{Hua-Xing Chen\inst{1}
\and Dan Zhou\inst{1}
\and Wei Chen\inst{2}
\thanks{wec053@mail.usask.ca}
\and Xiang Liu\inst{3,4}
\thanks{xiangliu@lzu.edu.cn}
\and Shi-Lin Zhu\inst{5,6,7}
\thanks{zhusl@pku.edu.cn}
}                     
\offprints{}          
\institute{
School of Physics and Beijing Key Laboratory of Advanced Nuclear Materials and Physics, Beihang University, Beijing 100191, China
\and
Department of Physics and Engineering Physics, University of Saskatchewan, Saskatoon, SK, S7N 5E2, Canada
\and
School of Physical Science and Technology, Lanzhou University, Lanzhou 730000, China
\and
Research Center for Hadron and CSR Physics, Lanzhou University and Institute of Modern Physics of CAS, Lanzhou 730000, China
\and
School of Physics and State Key Laboratory of Nuclear Physics and Technology, Peking University, Beijing 100871, China
\and
Collaborative Innovation Center of Quantum Matter, Beijing 100871, China
\and
Center of High Energy Physics, Peking University, Beijing 100871, China}
\date{Received: date / Revised version: date}
%
\abstract{
We give an explicit QCD sum rule investigation to hidden-charm baryonium states with the quark content $u\bar u d\bar d c\bar c$, spin $J=0/1/2/3$, and of both positive and negative parities. We systematically construct the relevant local hidden-charm baryonium interpolating currents, which can actually couple to various structures, including hidden-charm baryonium states, charmonium states plus two pions, and hidden-charm tetraquark states plus one pion, etc. We do not know which structure these currents couple to at the beginning, but after sum rule analyses we can obtain some information. We find some of them can couple to hidden-charm baryonium states, using which we evaluate the masses of the lowest-lying hidden-charm baryonium states with quantum numbers $J^P=2^-/3^-/0^+/1^+/2^+$ to be around 5.0 GeV. We suggest to search for hidden-charm baryonium states, especially the one of $J=3^-$, in the $D$-wave $J/\psi \pi \pi$ and $P$-wave $J/\psi \rho$ and $J/\psi \omega$ channels in this energy region.
} 
\maketitle

\section{Introduction}
\label{sec:intro}

Exploring exotic matter beyond conventional quark model is one of the most intriguing current research topics of hadronic physics.
With significant experimental progress on this issue over the past decade, dozens of charmonium/bottomonium-like $XYZ$ states were reported~\cite{Agashe:2014kda}, which are hidden-charm/bottom tetraquark candidates.
Besides them, the $P_c(4380)$ and $P_c(4450)$ were observed in the LHCb experiment in 2015~\cite{Aaij:2015tga}, which are hidden-charm pentaquark candidates.
They are new blocks of QCD matter, providing important hints to deepen our understanding of the non-perturbative QCD~\cite{Brambilla:2010cs,Liu:2013waa,Chen:2016qju}.
Facing their observations, we naturally conjecture that there should exist hidden-charm hexaquark states, and now is the time to hunt for them~\cite{Li:2014gra,Karliner:2015ina}.

In this letter we study one kind of hidden-charm hexaquark states, that is the hidden-charm baryonium states consisting of color-singlet heavy baryons and antibaryons. There have been several references on this issue, where
the $Y(4260)$, $Y(4360)$, $Y(4630)$, etc. were interpreted as hidden-charm baryonium states~\cite{Qiao:2005av,Qiao:2007ce,Lee:2011rka,Chen:2011cta,Meguro:2011nr,Li:2012bt,Chen:2013sba}.
However, these states can also be interpreted as hidden-charm tetraquark states~\cite{Agashe:2014kda,Chen:2016qju}, so better hidden-charm hexaquark candidates are still waiting to be found.

In this letter we perform an explicit QCD sum rule investigation to hidden-charm baryonium states with the quark content $u\bar u d\bar d c\bar c$, spin $J=0/1/2/3$, and of both positive and negative parities. We systematically construct the relevant local hidden-charm baryonium interpolating currents in Sec.~\ref{sec:current}. We find these currents can couple to various structures, including hidden-charm baryonium states [$\bar cc\bar qq \bar qq$], charmonium states plus two pions [$\bar cc + \pi\pi$], and hidden-charm tetraquark states plus one pion [$\bar cc\bar qq$+$\pi$], etc., which fact can be helpful to relevant studies, such as Lattice QCD.
We do not know which structure these currents couple to at the beginning, but we try to perform sum rule analyses using them in Sec.~\ref{sec:sumrule} to obtain some information. We find some of them can couple to hidden-charm baryonium states, using which we evaluate the masses of the lowest-lying hidden-charm baryonium states to be around 5.0 GeV, significantly larger than the masses of hidden-charm tetraquark states. Accordingly, we suggest to search for hidden-charm baryonium states, especially the one of $J=3^-$, in the $D$-wave $J/\psi \pi \pi$ and $P$-wave $J/\psi \rho$ and $J/\psi \omega$ channels in this energy region. The results are discussed and summarized in Sec.~\ref{sec:summary}.
This paper has a supplementary file ``OPE.nb'' containing all the spectral densities.

\section{Hidden-charm baryonium currents}
\label{sec:current}

To start our study, we first systematically construct local hidden-charm baryonium interpolating currents with the quark content $u\bar u d\bar d c\bar c$, spin $J=0/1/2/3$, and of both positive and negative parities. We use $\eta_{\mu_1\cdots\mu_j}$ to denote the current of spin $J$, and assume it couples to the physical state $X_j$ of spin $J$ through
\begin{eqnarray}
\langle 0 | \eta_{\mu_1\cdots\mu_j} | X_{j} \rangle &=& f_X \epsilon_{\mu_1\cdots\mu_j} \, ,
\label{eq:coupling}
\end{eqnarray}
where $f_X$ is the decay constant, and $\epsilon_{\mu_1\cdots\mu_j}$ is the traceless and symmetric polarization tensor.

There are two possible color configurations. One has a baryonium structure $[\epsilon^{abc}u_a d_b c_c][\epsilon^{def}\bar u_d \bar d_e \bar c_f]$, where $a \cdots f$ are color indices; the other has a three-meson structure, such as $[\bar c_a c_a][\bar u_b u_b][\bar d_c d_c]$, etc. The former configuration can be transformed to the latter through the the color rearrangement:
\begin{eqnarray}
\epsilon^{abc} \epsilon^{def} &=& \delta^{ad} \delta^{be} \delta^{cf} - \delta^{ae} \delta^{bd} \delta^{cf} + \delta^{af} \delta^{ce} \delta^{bd}
\label{eq:cr}
\\ \nonumber && ~~~ - \delta^{ad} \delta^{bf} \delta^{ce} + \delta^{ae} \delta^{bf} \delta^{cd} - \delta^{af} \delta^{cd} \delta^{be} \, .
\end{eqnarray}
We note that the latter can not be transformed back, suggesting that the three-meson structure is more complicated than the baryonium structure.
Eq.~(\ref{eq:cr}) implies that the hidden-charm baryonium currents can also couple to charmonium states plus two pions, such as $\eta_c \pi \pi$ and $J/\psi \pi \pi$, through
\begin{eqnarray}
\nonumber \langle 0 | \eta_{\mu_1\cdots\mu_j} | [\bar c c + \pi \pi] \rangle &=& f_{3M} \times \cdots \, ,
\end{eqnarray}
where $\cdots$ denotes relevant polarization tensors. Moreover, the hidden-charm baryonium currents may also couple to the $XYZ$ charmonium-like states (taken as hidden-charm tetraquark states) plus one pion through
\begin{eqnarray}
\nonumber \langle 0 | \eta_{\mu_1\cdots\mu_j} | [\bar c c \bar q q + \pi] \rangle &=& f_{2M} \times \cdots \, .
\end{eqnarray}
We shall find that both of these two cases are possible. Besides them, the diquark-anti-diquark-meson structure is also possible. At the beginning we do not know which structure the current $J$ couples to, but after sum rule analyses we can obtain some information.

To construct local baryonium interpolating currents, we first systematically construct local heavy baryon fields with the quark content $ud c$. We only investigate the fields of the following type
\begin{eqnarray}
[\epsilon^{abc} (u^T_a C \Gamma_i d_b) \Gamma_j c_c] \, ,
\end{eqnarray}
where $\Gamma_{i,j}$ are various Dirac matrices. The fields of the other types $[\epsilon^{abc} (u_a^T C \Gamma_i c_b) \Gamma_j d_c]$ and $[\epsilon^{abc} (d_a^T C \Gamma_i c_b) \Gamma_j u_c]$, etc. can be related to these fields through the Fierz transformation.

The $u$ and $d$ quarks can be either flavor symmetric or antisymmetric, and we use the $SU(3)$ flavor representations $\mathbf{6}_F$ and $\mathbf{\bar 3}_F$ to denote them, respectively. We can easily construct them based on the results of Ref.~\cite{Chen:2008qv}:
\begin{enumerate}

\item
We use $\mathcal{B}$ to denote the Dirac baryon fields without free Lorentz indices. There are
three fields of the flavor $\mathbf{\bar 3}_F$:
\begin{eqnarray}
\nonumber \Lambda_1 &=& \epsilon^{abc} (u_a^T C d_b) \gamma_5 c_c \, ,
\\ \Lambda_2 &=& \epsilon^{abc} (u_a^T C \gamma_5 d_b) c_c \, ,
\label{def:3D}
\\ \nonumber \Lambda_3 &=& \epsilon^{abc} (u_a^T C \gamma^\alpha \gamma_5 d_b) \gamma_\alpha c_c \, ,
\end{eqnarray}
and two fields of the flavor $\mathbf{6}_F$:
\begin{eqnarray}
\Sigma_1 &=& \epsilon^{abc} (u_a^T C \gamma^\alpha d_b) \gamma_\alpha \gamma_5 c_c \, ,
\label{def:6D}
\\ \nonumber \Sigma_2 &=& \epsilon^{abc} (u_a^T C \sigma^{\alpha\beta} d_b) \sigma_{\alpha\beta} \gamma_5 c_c \, .
\end{eqnarray}
All these fields $\mathcal{B}$ have the spin-parity $J^P=1/2^+$, while $\gamma_5 \mathcal{B}$ have $1/2^-$.

\item
We use $\mathcal{B}_\mu$ to denote the baryon fields with one free Lorentz index. There is one field of the flavor $\mathbf{\bar 3}_F$:
\begin{eqnarray}
\Lambda_{4\mu} &=& \epsilon^{abc} (u_a^T C \gamma_\mu \gamma_5 d_b) \gamma_5 c_c \, ,
\label{def:3R}
\end{eqnarray}
and two fields of the flavor $\mathbf{6}_F$:
\begin{eqnarray}
\Sigma_{3\mu} &=& \epsilon^{abc} (u^{aT} C \gamma_\mu d^b) c^c \, ,
\label{def:6R}
\\ \nonumber \Sigma_{4\mu} &=& \epsilon_{abc} (u^{aT} C \sigma_{\mu\alpha} d^b) \gamma^{\alpha} c^c \, .
\end{eqnarray}
All these fields contain the spin-parity $J^P=3/2^+$ components, while $\gamma_5 \mathcal{B}_\mu$ have $3/2^-$.

\item
We use $\mathcal{B}_{\mu\nu}$ to denote the baryon fields with two free antisymmetric Lorentz indices. There is only one field of the flavor $\mathbf{6}_F$:
\begin{eqnarray}
\Sigma_{5\mu\nu} &=& \epsilon_{abc} (u^{aT} C \sigma_{\mu\nu} d^b) \gamma_5 c^c \, .
\label{def:6T}
\end{eqnarray}
This field also contains the spin $J=3/2$ component, but it contains both positive parity and negative parity components.

\end{enumerate}

Based on these heavy baryon fields, the local hidden-charm baryonium interpolating currents with the color configuration
$[\epsilon^{abc}u_a d_b c_c][\epsilon^{def}\bar u_d \bar d_e \bar c_f]$ can be systematically
constructed:
\begin{enumerate}

\item
We find altogether 44 baryonium currents of spin $J=0$. Half of these currents
have the positive parity:
\begin{eqnarray}
\overline{\mathcal{B}} \mathcal{B} \, , \,
\overline{\mathcal{B}}_\alpha \mathcal{B}^\alpha \, , \,
\overline{\mathcal{B}}_{\alpha\beta} \mathcal{B}^{\alpha\beta} \, , \,
\end{eqnarray}
and the other half have the negative parity:
\begin{eqnarray}
\overline{\mathcal{B}} \gamma_5 \mathcal{B} \, , \,
\overline{\mathcal{B}}_\alpha \gamma_5 \mathcal{B}^\alpha \, , \,
\overline{\mathcal{B}}_{\alpha\beta} \gamma_5 \mathcal{B}^{\alpha\beta} \, . \,
\end{eqnarray}
Other currents, such as $\overline{\mathcal{B}}_\alpha \gamma^\alpha \mathcal{B}$, $\overline{\mathcal{B}}_\alpha \sigma^{\alpha\beta} \mathcal{B}_\beta$, etc.
can be related to these currents.

\item
We find altogether 80 baryonium currents of spin $J=1$. Half of these currents
have the positive parity:
\begin{eqnarray}
&
\overline{\mathcal{B}} \gamma_\mu \gamma_5 \mathcal{B} \, , \,
\overline{\mathcal{B}}_\alpha \gamma_\mu \gamma_5 \mathcal{B}^\alpha \, , \,
\overline{\mathcal{B}}_{\alpha\beta} \gamma_\mu \gamma_5 \mathcal{B}^{\alpha\beta} \, , \,
&
\\ \nonumber
&
\overline{\mathcal{B}} \mathcal{B}_\mu \, , \,
\overline{\mathcal{B}}^\alpha \mathcal{B}_{\alpha\mu} \, , \,
&
\end{eqnarray}
and the other half have the negative parity:
\begin{eqnarray}
&
\overline{\mathcal{B}} \gamma_\mu \mathcal{B} \, , \,
\overline{\mathcal{B}}_\alpha \gamma_\mu \mathcal{B}^\alpha \, , \,
\overline{\mathcal{B}}_{\alpha\beta} \gamma_\mu \mathcal{B}^{\alpha\beta} \, , \,
&
\\ \nonumber
&
\overline{\mathcal{B}} \gamma_5 \mathcal{B}_\mu \, , \,
\overline{\mathcal{B}}^\alpha \gamma_5 \mathcal{B}_{\alpha\mu} \, . \,
&
\end{eqnarray}
Other currents, such as $\overline{\mathcal{B}}^\alpha \sigma_{\mu\alpha} \mathcal{B}$, $\overline{\mathcal{B}}_{\mu\alpha} \gamma^\alpha \mathcal{B}$, etc.
can be related to these currents.

\item
We find altogether 50 baryonium currents of spin $J=2$. Half of these currents
have the positive parity:
\begin{eqnarray}
&
\mathcal{S}_2 \big[ \overline{\mathcal{B}}_{\mu_1} \mathcal{B}_{\mu_2} \big] \, , \,
\mathcal{S}_2 \big[ \overline{\mathcal{B}}_{\mu_1\alpha} \mathcal{B}_{\mu_2\alpha} \big] \, , \,
&
\\ \nonumber
&
\mathcal{S}_2 \big[ \overline{\mathcal{B}} \gamma_{\mu_1} \gamma_5 \mathcal{B}_{\mu_2} \big] \, , \,
\mathcal{S}_2 \big[ \overline{\mathcal{B}}^\alpha \gamma_{\mu_1} \gamma_5 \mathcal{B}_{\mu_2\alpha} \big] \, . \,
&
\end{eqnarray}
where $\mathcal{S}_j$ denotes symmetrization in the sets $(\mu_1 \cdots \mu_j)$. The other half have the negative parity:
\begin{eqnarray}
&
\mathcal{S}_2 \big[ \overline{\mathcal{B}}_{\mu_1} \gamma_5 \mathcal{B}_{\mu_2} \big] \, , \,
\mathcal{S}_2 \big[ \overline{\mathcal{B}}_{\mu_1\alpha} \gamma_5 \mathcal{B}_{\mu_2\alpha} \big] \, , \,
&
\\ \nonumber
&
\mathcal{S}_2 \big[ \overline{\mathcal{B}} \gamma_{\mu_1} \mathcal{B}_{\mu_2} \big] \, , \,
\mathcal{S}_2 \big[ \overline{\mathcal{B}}^\alpha \gamma_{\mu_1} \mathcal{B}_{\mu_2\alpha} \big] \, . \,
&
\end{eqnarray}
Other currents, such as $\mathcal{S}_2 \big[ \overline{\mathcal{B}}_\mu \gamma^\alpha \mathcal{B}_{\nu\alpha} \big]$, etc.
can be related to these currents.

\item
We find altogether 14 baryonium currents of spin $J=3$. Half of these currents
have the positive parity:
\begin{eqnarray}
&
\mathcal{S}_3 \big[ \overline{\mathcal{B}}_{\mu_1} \gamma_{\mu_2} \gamma_5 \mathcal{B}_{\mu_3} \big] \, , \,
\mathcal{S}_3 \big[ \overline{\mathcal{B}}_{\mu_1\alpha} \gamma_{\mu_2} \gamma_5 \mathcal{B}_{\mu_3\alpha} \big] \, , \,
&
\end{eqnarray}
and the other half have the negative parity:
\begin{eqnarray}
&
\mathcal{S}_3 \big[ \overline{\mathcal{B}}_{\mu_1} \gamma_{\mu_2}5 \mathcal{B}_{\mu_3} \big] \, , \,
\mathcal{S}_3 \big[ \overline{\mathcal{B}}_{\mu_1\alpha} \gamma_{\mu_2} \mathcal{B}_{\mu_3\alpha} \big] \, , \,
&
\end{eqnarray}
Other currents, such as $\mathcal{S}_3 \big[ \overline{\mathcal{B}}_{\mu_1\alpha} \sigma_{\mu_2\alpha} \mathcal{B}_{\mu_3} \big]$, etc.
can be related to these currents.

\end{enumerate}
We can also construct some other baryonium currents which contain two free antisymmetric Lorentz indices, such as $\overline{\mathcal{B}} \mathcal{B}_{\mu\nu}$, $\overline{\mathcal{B}} \sigma_{\mu\nu} \mathcal{B}_\rho$, $\overline{\mathcal{B}} \sigma_{\mu\nu} \mathcal{B}_{\rho\sigma}$, etc.
However, these currents contain both positive parity and negative parity components, which we shall not investigate in the present study for simplicity.

\section{QCD sum rule analyses}
\label{sec:sumrule}

In the following, we use the method of QCD sum rules~\cite{Shifman:1978bx,Reinders:1984sr,Nielsen:2009uh,Chen:2015moa} to investigate these hidden-charm baryonium currents,
and try to find out which structure these currents couple to.
The two-point correlation function can be written as:
\begin{eqnarray}
\label{eq:pi} && \Pi_{\mu_1 \cdots \mu_j,\nu_1 \cdots \nu_j}\left(q^2\right)
\\ \nonumber &=& i \int d^4x e^{iq\cdot x} \langle 0 | T\left[\eta_{\mu_1 \cdots \mu_j}(x) \eta^\dagger_{\nu_1 \cdots \nu_j}(0)\right] | 0 \rangle
\\ \nonumber &=& (-1)^j \mathcal{S}^\prime_j
\left[g_{\mu_1 \nu_1} \cdots g_{\mu_j \nu_j}\right] \Pi_j \left(q^2\right) + \cdots \, ,
\end{eqnarray}
where $\mathcal{S}^\prime_j$ denotes symmetrization and subtracting the trace
terms in the sets $(\mu_1 \cdots \mu_j)$ and $(\nu_1 \cdots \nu_j)$.
Although the current $\eta_{\mu_1 \cdots \mu_j}$ contains all spin $0$--$j$ components,
the leading term $\Pi_j \left(q^2\right)$ has been totally symmetrized and only contains the spin $j$ component, while $\cdots$ contains other spin components.
Hence, we shall only calculate $\Pi_j \left(q^2\right)$ and use it to perform sum rule analyses.

We can calculate the two-point correlation function Eq.~(\ref{eq:pi}) in the QCD operator product expansion (OPE) up to certain order in the expansion,
which is then matched with a hadronic parametrization to extract information about hadron properties. Following Ref.~\cite{Chen:2015moa},
we obtain the mass of $X_{j}$ to be
%
\begin{eqnarray}
M^2_X(s_0, M_B) 
&=& {\int^{s_0}_{s_<} e^{-s/M_B^2} \rho(s) s ds \over \int^{s_0}_{s_<} e^{-s/M_B^2} \rho(s) ds} \, ,
\label{eq:mass}
\end{eqnarray}
%
where $\rho(s)$ is the QCD spectral density which we evaluate up to dimension twelve, including the perturbative term, the quark condensate $\langle \bar q q \rangle$, the gluon condensate $\langle g_s^2 GG \rangle$, the quark-gluon mixed condensate $\langle g_s \bar q \sigma G q \rangle$, and their combinations $\langle \bar q q \rangle^2$,
$\langle \bar q q \rangle\langle g_s \bar q \sigma G q \rangle$, $\langle g_s \bar q \sigma G q \rangle^2$ and $\langle \bar q q \rangle^4$. The full expressions are lengthy and listed in the supplementary file ``OPE.nb''.
We use the values listed in Ref.~\cite{Chen:2015ata} for these condensates and the charm quark mass (see also Refs.~\cite{Yang:1993bp,Agashe:2014kda,Eidemuller:2000rc,Narison:2002pw,Gimenez:2005nt,Jamin:2002ev,Ioffe:2002be,Ovchinnikov:1988gk,colangelo}).

There are two free parameters in Eq.~(\ref{eq:mass}): the Borel mass $M_B$ and the threshold value $s_0$.
As the first step, we fix $M_B^2 = 4.0$ GeV$^2$, and investigate the $s_0$ dependence. We find that all
the currents can be classified into six types, except two currents whose OPE can not be easily
calculated (they are $\bar \Sigma_{2}\gamma_\mu\Sigma_{2}$ and $\bar \Sigma_{2}\gamma_\mu\gamma_5\Sigma_{2}$ of $J^P=1^-$ and $1^+$ respectively):
\begin{enumerate}

\item[A.] [Non-Structure] Many currents seemingly do not couple to any structure, in other words, these currents seem to well couple to the continuum.
Take the current $\bar \Lambda_1 \gamma_5 \Lambda_2$ of $J^P = 0^-$ as an example. We use it to perform QCD sum rule analysis,
and show the obtained mass $M_X$ as a function of $s_0$ in the left panel of Fig.~\ref{fig:typeA}. We find that $M_X$
quickly and monotonically increases with $s_0$, suggesting that this current does not couple to any structure.
Sometimes the mass curves behave differently, as shown in the middle and right panels of Fig.~\ref{fig:typeA}, where
the currents $\bar \Sigma_{2} \gamma_5 \Sigma_{3\mu}$ and $\bar \Sigma_{5\nu} \gamma_\mu \Sigma_{5}^\nu$ both of $J^P = 1^-$ are used as examples. However, still no structure can be clearly verified.

\begin{figure*}[hbt]
\begin{center}
\scalebox{0.465}{\includegraphics{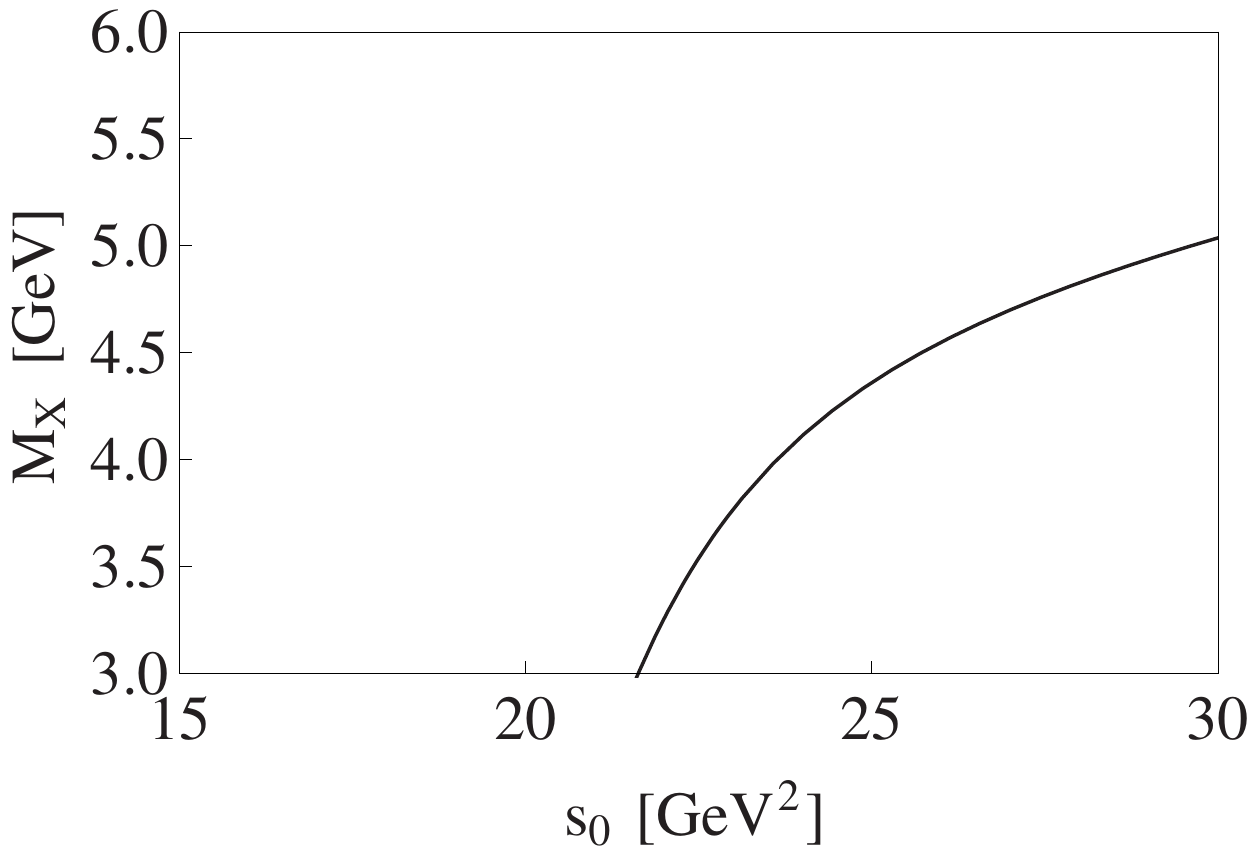}}
\scalebox{0.465}{\includegraphics{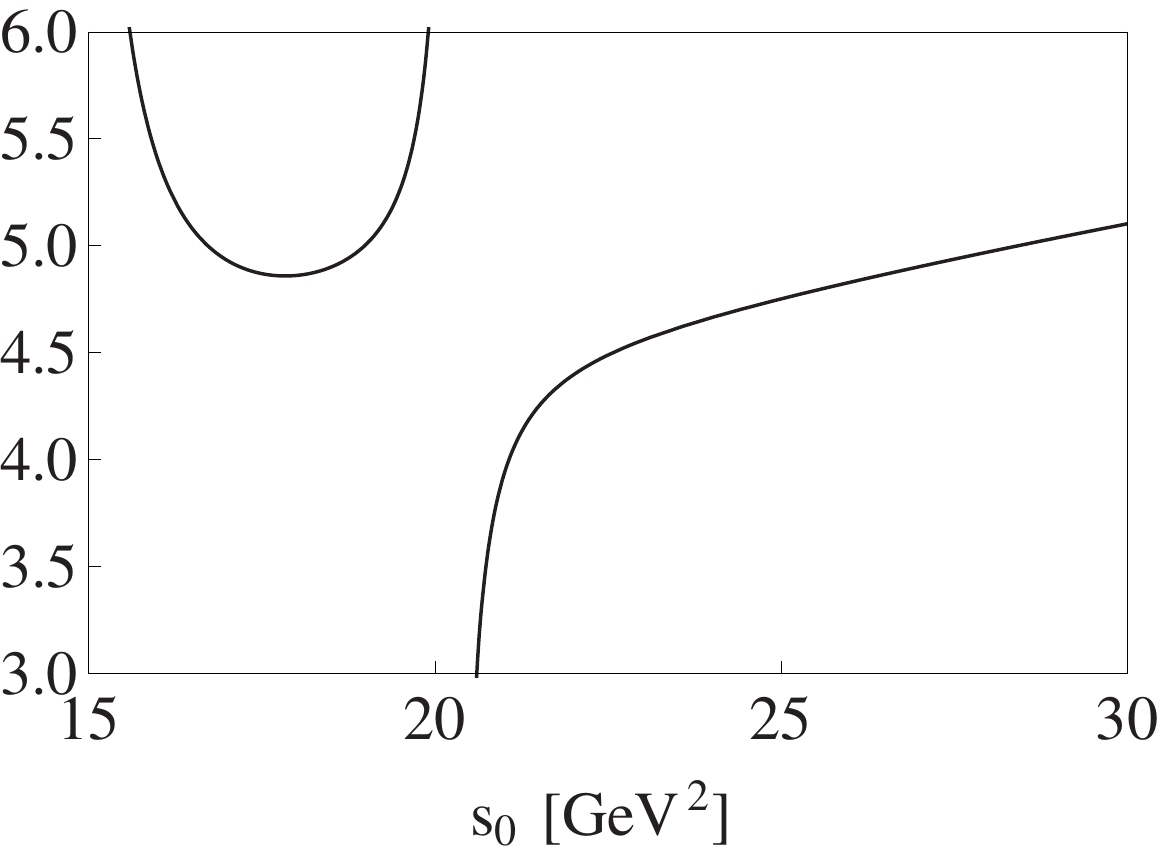}}
\scalebox{0.485}{\includegraphics{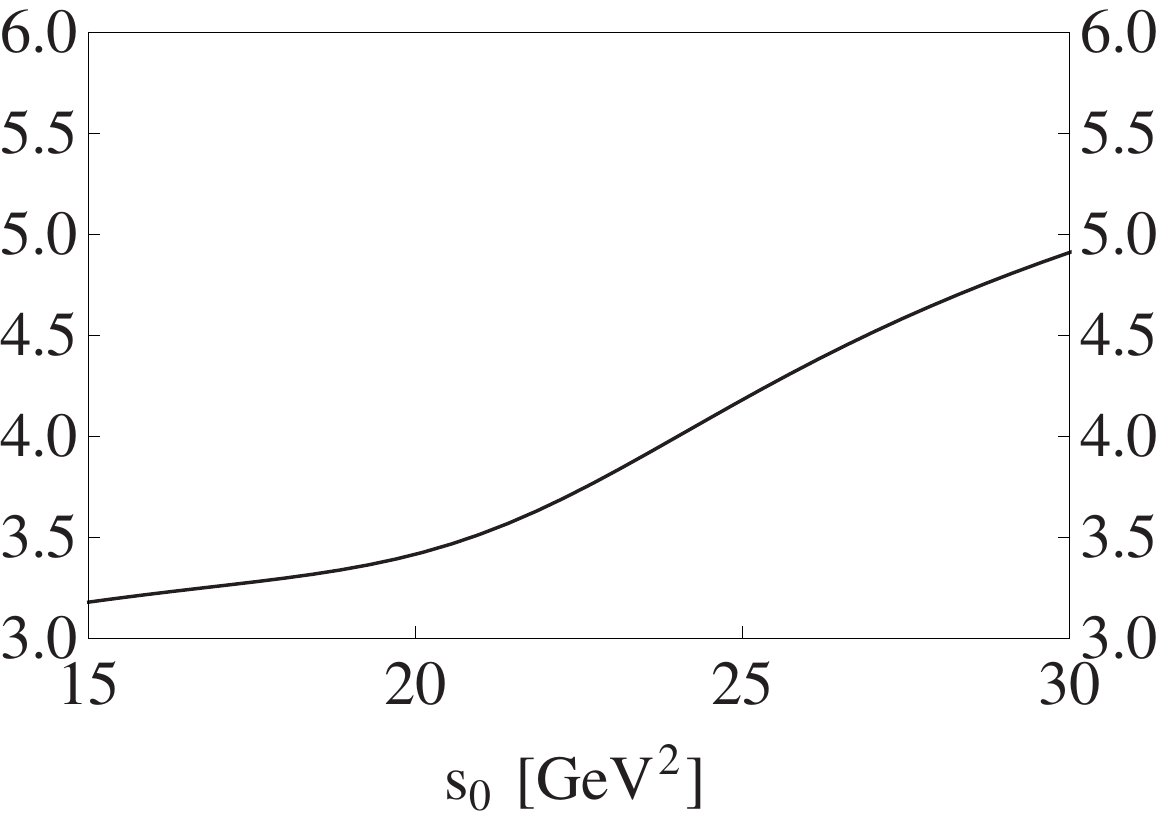}}
\caption{Type A [Non-Structure]: Variations of $M_X$ with respect to the threshold value $s_0$, when the Borel mass $M_B$ is fixed to be $M_B^2 = 4.0$ GeV$^2$.
The currents $\bar \Lambda_1 \gamma_5 \Lambda_2$ (left), $\bar \Sigma_{2} \gamma_5 \Sigma_{3\mu}$ (middle), and $\bar \Sigma_{5\nu} \gamma_\mu \Sigma_{5}^\nu$ (right) are used as examples.}
\label{fig:typeA}
\end{center}
\end{figure*}

\item[B.] [$\bar cc + \pi\pi$] Many currents seem to well couple to charmonium states plus two pions.
Take the current $\bar \Sigma_1 \gamma_5 \Lambda_1$ of $J^P = 0^-$ as an example. Its obtained mass $M_X$
is shown as a function of $s_0$ in the left panel of Fig.~\ref{fig:typeB}. There is a mass plateau around
3.5 GeV in a wide region of 15 GeV$^2 < s_0 < 25$ GeV$^2$, suggesting that this current well couple to charmonium states
plus two pions. Take the current $\bar \Sigma_2 \gamma_5 \Lambda_1$ of $J^P = 0^-$ as another example. Its obtained mass $M_X$
is shown as a function of $s_0$ in the right panel of Fig.~\ref{fig:typeB}. $M_X$ is around
3.5 GeV in a narrower region of 15 GeV$^2 < s_0 < 20$ GeV$^2$, still suggesting that this current couple to charmonium states
plus two pions.

\begin{figure*}[hbt]
\begin{center}
\scalebox{0.65}{\includegraphics{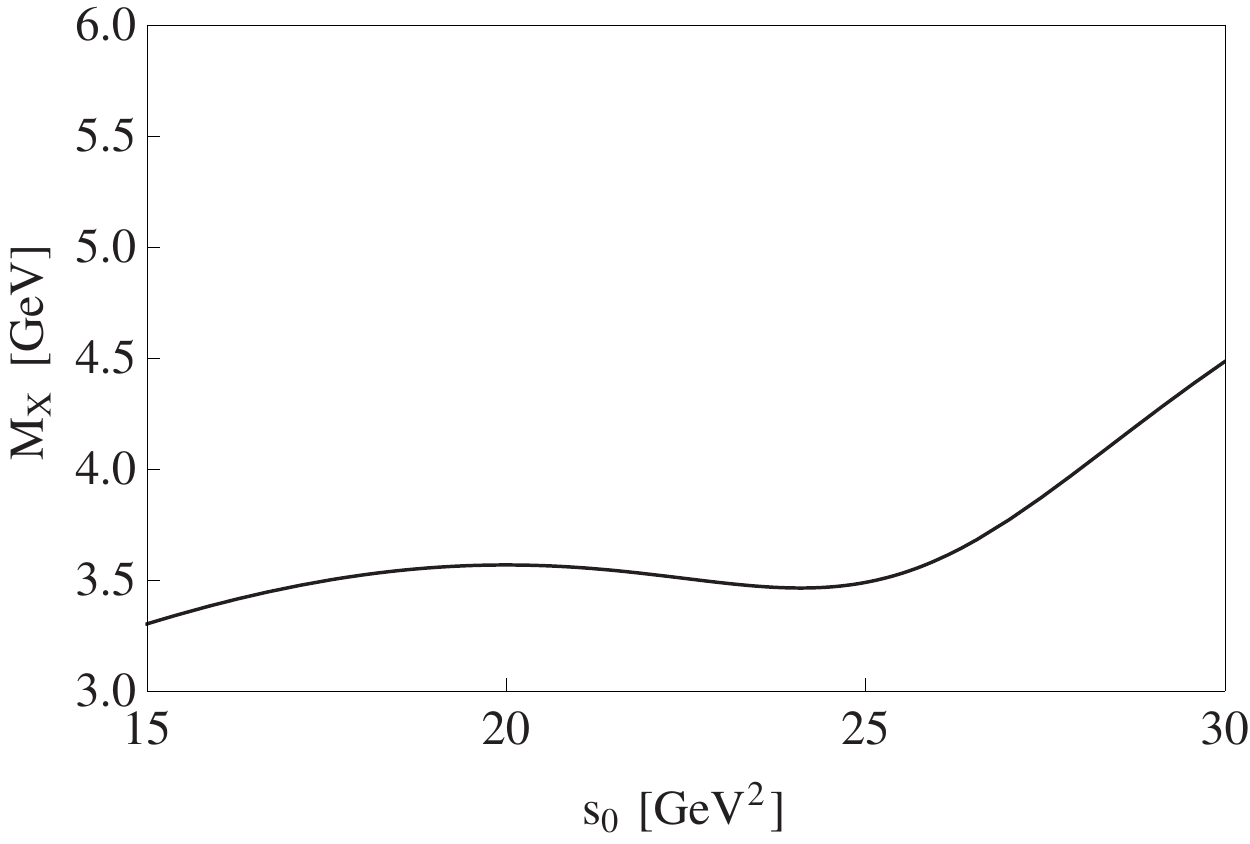}}
\scalebox{0.673}{\includegraphics{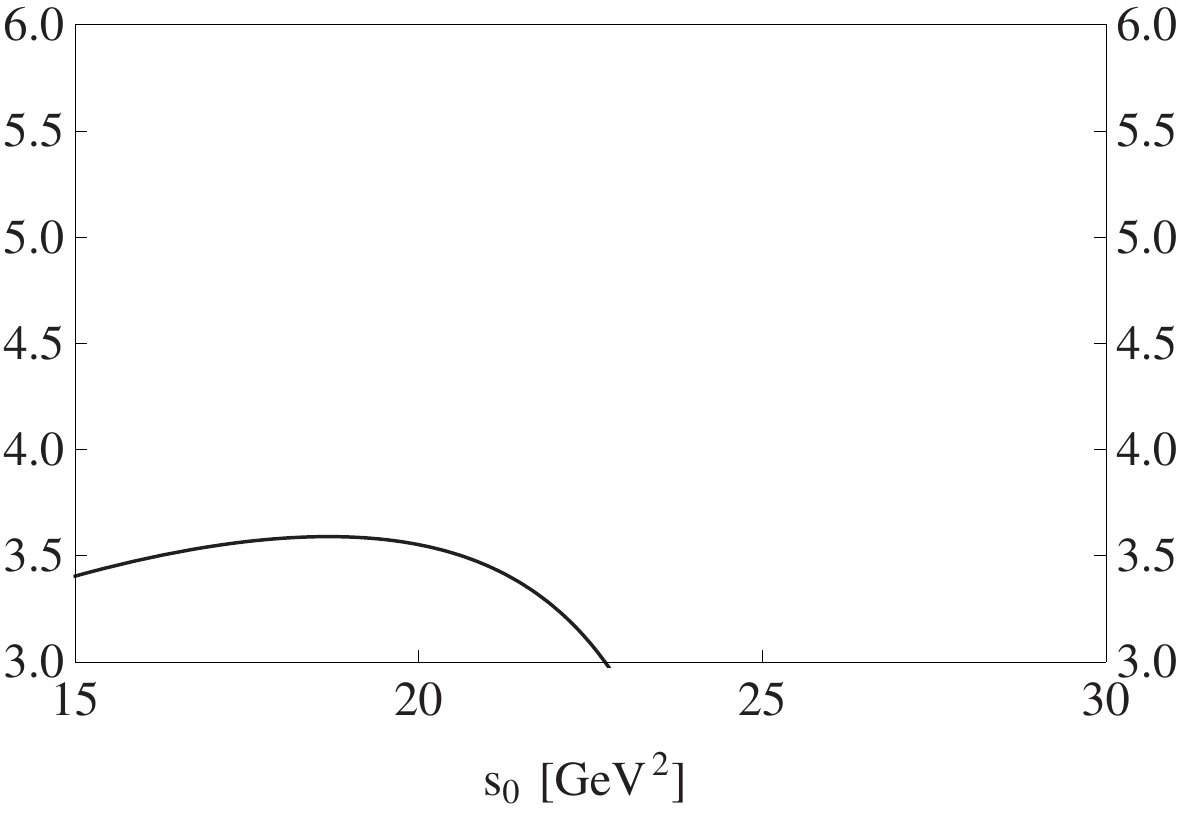}}
\caption{Type B [$\bar cc$+$\pi\pi$]: Variations of $M_X$ with respect to the threshold value $s_0$, when the Borel mass $M_B$ is fixed to be $M_B^2 = 4.0$ GeV$^2$.
The currents $\bar \Sigma_1 \gamma_5 \Lambda_1$ (left) and $\bar \Sigma_2 \gamma_5 \Lambda_1$ (right) are used as examples.}
\label{fig:typeB}
\end{center}
\end{figure*}

\item[C.] [$\bar cc\bar qq$+$\pi$] Several currents seem to couple to hidden-charm tetraquark states plus one pion.
Take the current $\mathcal{S}_2 \big[\bar \Lambda_{4\mu_1} \Lambda_{4\mu_2}\big]$ of $J^P = 2^+$ as an example. Its obtained mass $M_X$ is shown as a function of $s_0$ in the left panel
of Fig.~\ref{fig:typeCD}. There is a mass plateau around 4.2 GeV in a very wide region of 16 GeV$^2 < s_0 < 25$ GeV$^2$, suggesting that this current well couple
to hidden-charm tetraquark states plus one pion.

We note that there may exist hidden-charm baryonium states in this energy region, for example,
the $Y(4630)$ was suggested to be a $\Lambda_c \bar \Lambda_c$ bound state in Ref.~\cite{Lee:2011rka}. However, because
hidden-charm baryonium states in this region can not be well differentiated from hidden-charm tetraquark states, we shall not
pay attention to this case in this paper, but try to find more significant hidden-charm baryonium signals.
It is just for simplicity that we use [$\bar cc\bar qq$+$\pi$] to denote this type.

\begin{figure*}[hbt]
\begin{center}
\scalebox{0.65}{\includegraphics{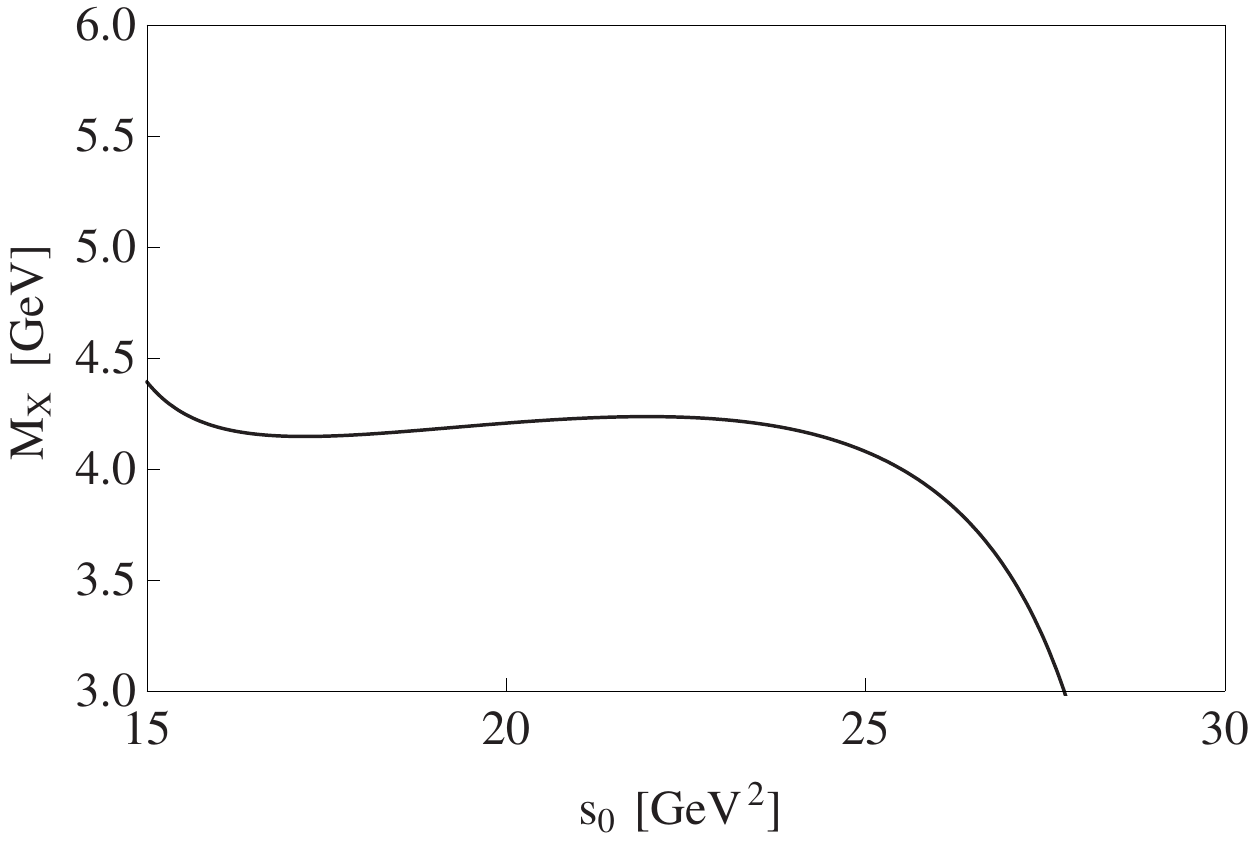}}
\scalebox{0.673}{\includegraphics{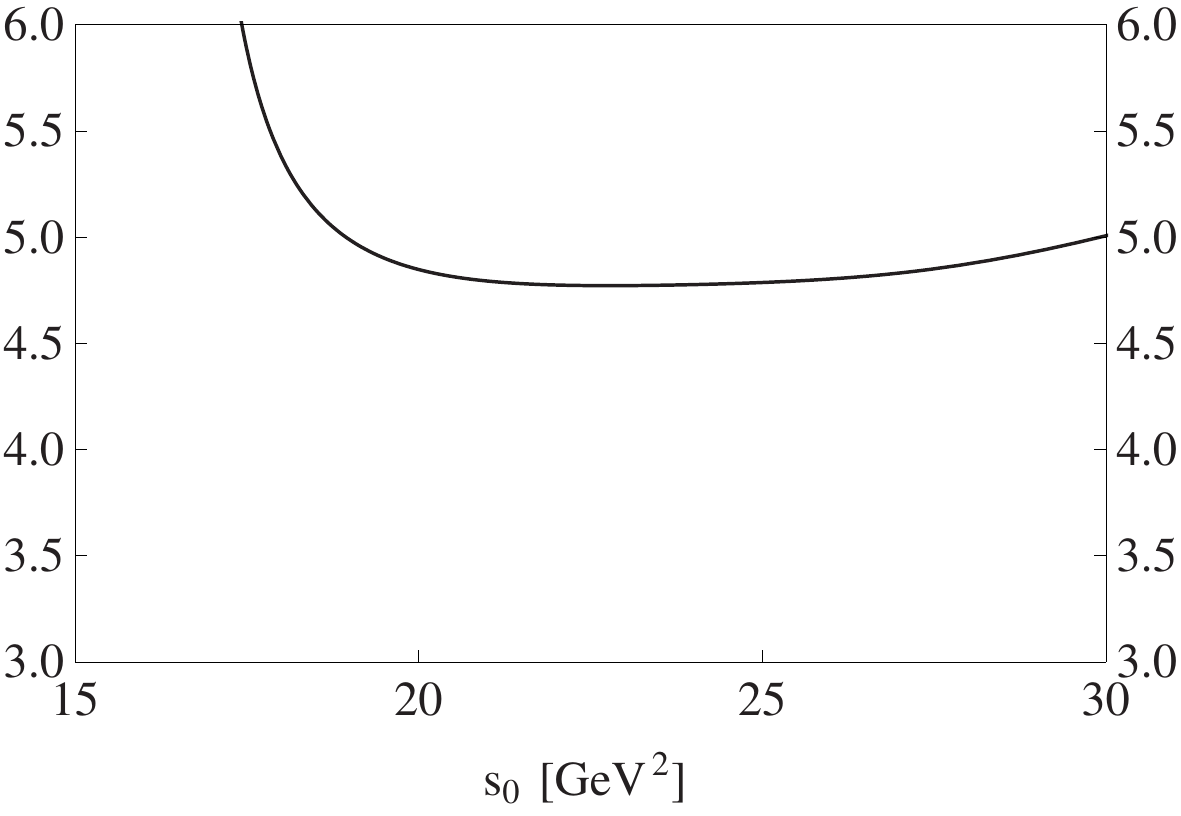}}
\caption{Variations of $M_X$ with respect to the threshold value $s_0$, when the Borel mass $M_B$ is fixed to be $M_B^2 = 4.0$ GeV$^2$.
Left: Type C [$\bar cc\bar qq$+$\pi$], where the current $\mathcal{S}_2 \big[\bar \Lambda_{4\mu_1} \Lambda_{4\mu_2}\big]$ is used as an example;
Right: Type D [$\bar cc\bar qq \bar qq$], where the current $\bar \Lambda_3 \Lambda_{4\mu}$ is used as an example.
}
\label{fig:typeCD}
\end{center}
\end{figure*}

\item[D.] [$\bar cc\bar qq \bar qq$] Many currents seem to well couple to hidden-charm baryonium states.
Take the current $\bar \Lambda_3 \Lambda_{4\mu}$ of $J^P = 1^+$ as an example. Its obtained mass $M_X$
is shown as a function of $s_0$ in the right panel of Fig.~\ref{fig:typeCD}. There is a mass plateau around 4.9 GeV
in a wide region of 20 GeV$^2 < s_0 < 30$ GeV$^2$. We shall use this type to detailly perform numerical analyses.

\item[E.] [Mixture of B and D] Many currents seem to couple to both charmonium states plus two pions and hidden-charm baryonium states.
Take the current $\bar \Lambda_{4\nu} \gamma_{\mu}\gamma_5 \Sigma_{3}^\nu$ of $J^P = 1^+$ as an example. Its obtained mass $M_X$ is shown as a function of $s_0$ in the middle panel of Fig.~\ref{fig:typeEF}.
We find two mass plateaus, one near 3.2 GeV and the other near 5.6 GeV, suggesting that this current may couple to both charmonium states plus two pions and hidden-charm baryonium states.
We note that the plateau in the lower-left corner is actually non-physical (the spectral density is negative in this region), but anyway we shall not pay attention to this case in this paper.

\begin{figure*}[hbt]
\begin{center}
\scalebox{0.675}{\includegraphics{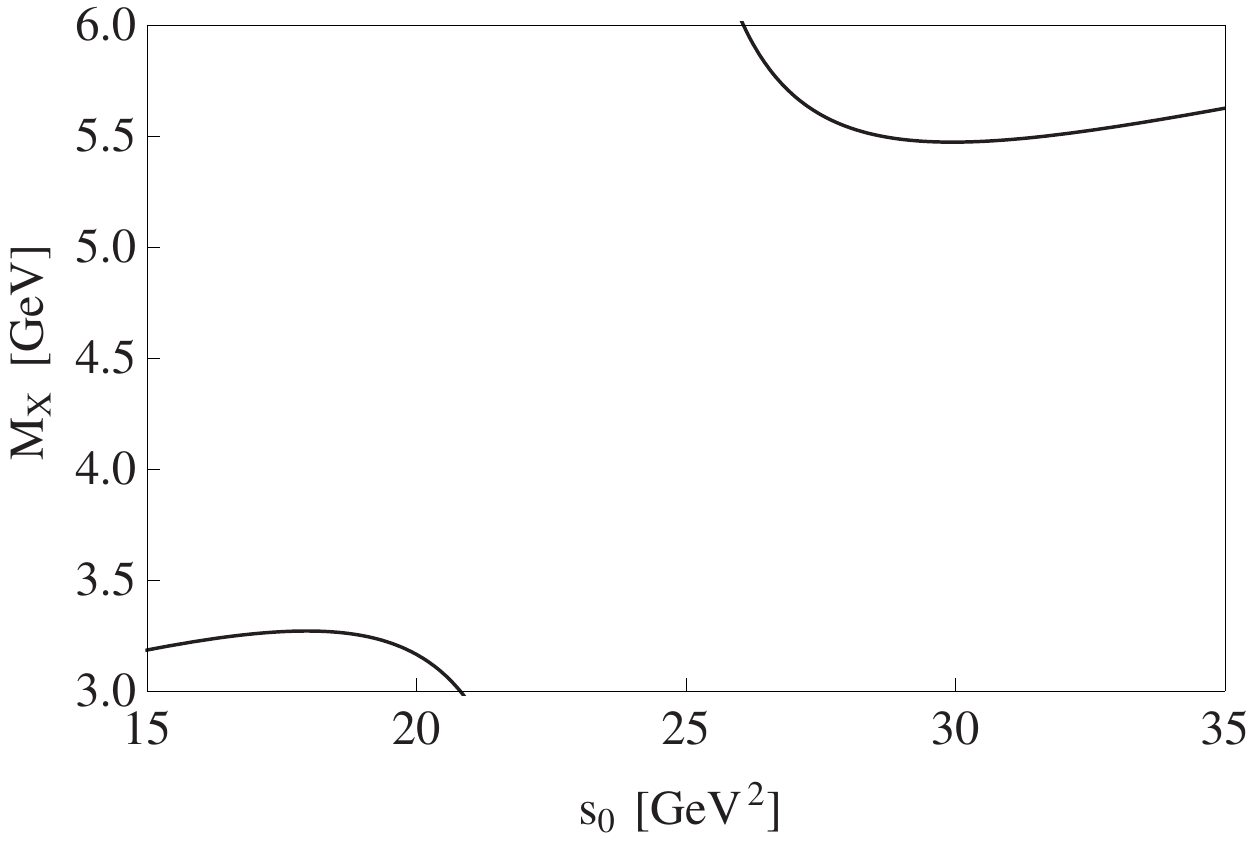}}
\scalebox{0.66}{\includegraphics{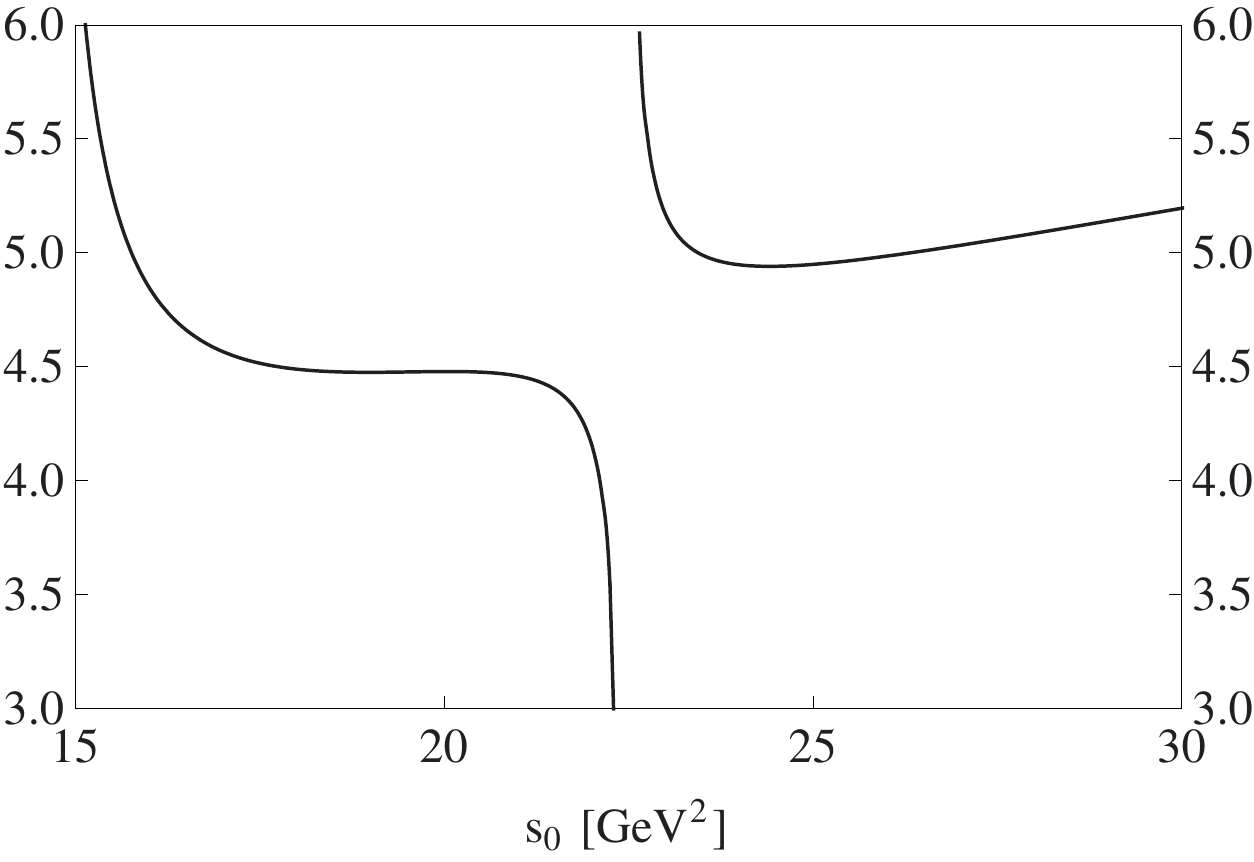}}
\caption{Variations of $M_X$ with respect to the threshold value $s_0$, when the Borel mass $M_B$ is fixed to be $M_B^2 = 4.0$ GeV$^2$.
Left: Type E [Mixture of B and D], where the current $\bar \Lambda_{4\nu} \gamma_{\mu}\gamma_5 \Sigma_{3}^\nu$ is used as an example;
Right: Type F [Mixture of C and D], where the current $\mathcal{S}_3 \big[\bar \Sigma_{4\mu_1} \gamma_{\mu_2} \Sigma_{3\mu_3}\big]$ is used as an example.
}
\label{fig:typeEF}
\end{center}
\end{figure*}

\item[F.] [Mixture of C and D] A few currents seem to couple to both hidden-charm tetraquark states plus one pion and hidden-charm baryonium states.
Take the current $\mathcal{S}_3 \big[\bar \Sigma_{4\mu_1} \gamma_{\mu_2} \Sigma_{3\mu_3}\big]$ of $J^P = 3^-$ as an example. Its obtained mass $M_X$ is shown as a function of $s_0$ in the right panel of Fig.~\ref{fig:typeEF}.
Again we find two mass plateaus (the left plateau is still non-physical), one near 4.5 GeV and the other near 5.0 GeV, suggesting that this current may couple to both hidden-charm tetraquark states plus one pion and hidden-charm baryonium states. We shall also use this type to detailly perform numerical analyses.

It is because of this type that we classified Type C to be hidden-charm tetraquark states plus one pion, i.e., it is less possible that this current couples to two different hidden-charm baryonium states, but more possible that it couples to two different structures, because it leads to two separated mass plateaus.

\end{enumerate}

\begin{table}[hbtp]
\begin{center}
\caption{Classification of hidden-charm baryonium currents. There are altogether six types:
Type A [Non-Structure], Type B [$\bar cc + \pi\pi$], Type C [$\bar cc\bar qq$+$\pi$], Type D [$\bar cc\bar qq \bar qq$], Type E [Mixture of B and D], and Type F [Mixture of C and D].
The numbers of currents belonging to these types are shown below.
There are two currents whose OPE can not be easily calculated. They are $\bar \Sigma_{2}\gamma_\mu\Sigma_{2}$ of $J^P=1^-$ and $\bar \Sigma_{2}\gamma_\mu\gamma_5\Sigma_{2}$ of $J^P=1^+$.
}
\begin{tabular}{c|cccc|cc|c}
\toprule[1pt]\toprule[1pt]
~~$J^{P}$~~ & ~~\mbox{A}~ & ~\mbox{B}~ & ~\mbox{C}~ & ~\mbox{D}~~ & ~~\mbox{E}~ & ~\mbox{F}~~ & ~~\mbox{Total}~~
\\ \midrule[1pt]
$0^-$ & 14 & 8  & 0 & 0  & 0  & 0 & 22
\\
$1^-$ & 23 & 16 & 0 & 0  & 0  & 0 & 39
\\
$2^-$ & 10 & 11 & 0 & 1  & 0  & 3 & 25
\\
$3^-$ & 2  & 3  & 0 & 0  & 0  & 2 & 7
\\ \midrule[1pt]
$0^+$ & 0  & 0  & 2 & 8  & 12 & 0 & 22
\\
$1^+$ & 0  & 2  & 2 & 14 & 21 & 0 & 39
\\
$2^+$ & 0  & 3  & 2 & 9  & 11 & 0 & 25
\\
$3^+$ & 0  & 2  & 1 & 1  & 3  & 0 & 7
\\ \bottomrule[1pt]\bottomrule[1pt]
\end{tabular}
\label{tab:number}
\end{center}
\end{table}

We show the numbers of currents belonging to these six types in Table~\ref{tab:number}, but note that sometimes the currents can not be well classified, for examples, some currents can be classified to both Type A and Type B (compare the right panel of Fig.~\ref{fig:typeA} and the left panel of Fig.~\ref{fig:typeB}). However, we need not solve this problem, because the currents of Type D are probably the best choices to study hidden-charm baryonium states.

\begin{figure*}[hbt]
\begin{center}
\scalebox{1}{\includegraphics{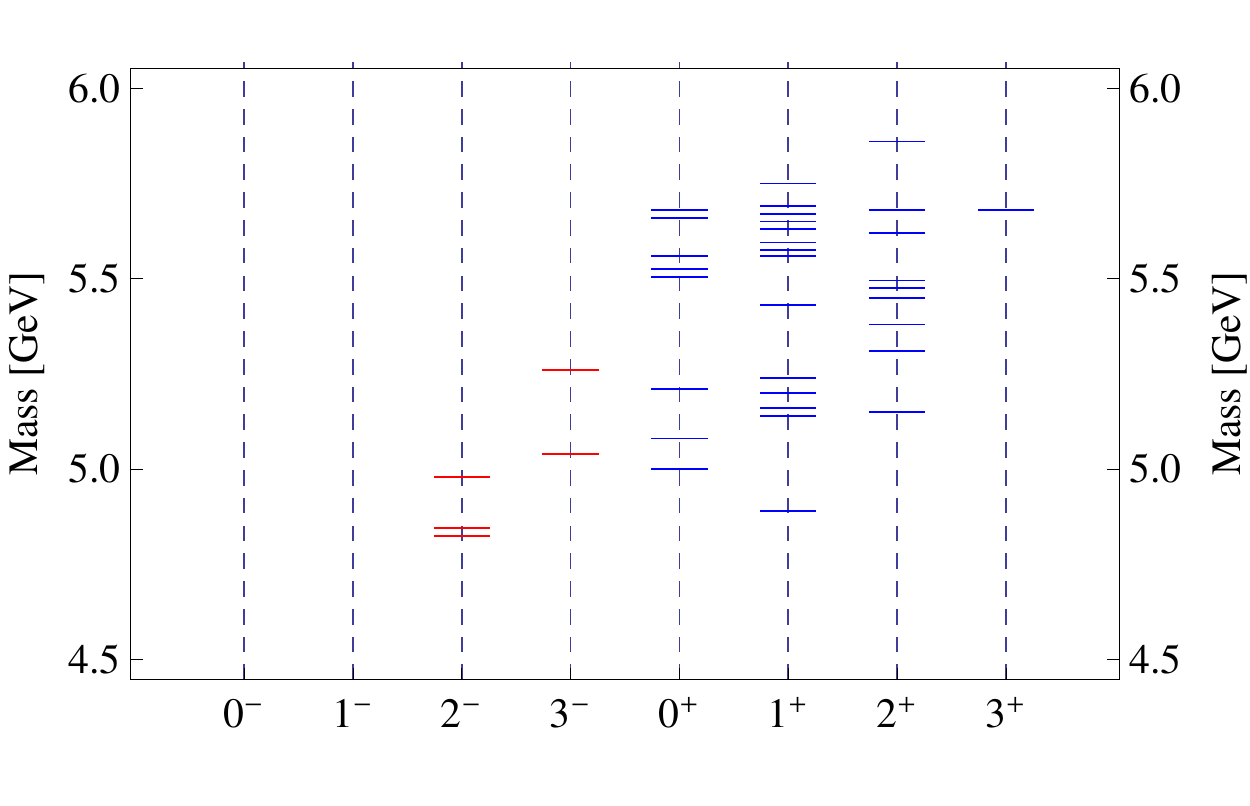}}
\caption{(Color online) Spectrum of hidden-charm baryonium states obtained using the method of QCD sum rules. The blue lines are obtained using the currents of Type D, and the red lines are obtained using the currents of Type F.}
\label{fig:spectrum}
\end{center}
\end{figure*}

Following Ref.~\cite{Chen:2015moa,Chen:2016otp}, we use the currents of Type D to perform numerical analyses and evaluate the masses of hidden-charm baryonium states. We list the results in Fig.~\ref{fig:spectrum} using blue lines, but leave the detailed analyses for our future studies. We note that the uncertainties of these results are about $\pm0.20$ GeV.
We find that all the currents of Type D with the positive parity lead to reasonable sum rules. However, the only current of Type D with the negative parity does not lead to a reasonable sum rule. Hence, we also use the currents of Type F with the negative parity to perform numerical analyses. These currents also lead to reasonable sum rules, whose results are listed in Fig.~\ref{fig:spectrum} using red lines.

We find that the masses of the lowest-lying hidden-charm baryonium states with quantum numbers $J^P=2^-/3^-/0^+/1^+/2^+$ are around 5.0 GeV. Especially, the one of $J^P = 3^-$ can not be accessed by the $S$-wave [$\bar cc + \pi\pi$] and [$\bar cc\bar qq$+$\pi$] structures, so it is a very good hidden-charm baryonium candidate for experimental observations. We evaluate its mass to be around 5.04 GeV (see the right panel of Fig.~4), and suggest to search for it in the LHCb and forthcoming BelleII experiments.

\section{Summary}
\label{sec:summary}

To summarize the current letter, we use the method of QCD sum rule to investigate hidden-charm baryonium states with the quark content $u\bar u d\bar d c\bar c$, spin $J=0/1/2/3$, and of both positive and negative parities. We systematically construct the relevant local hidden-charm baryonium interpolating currents. We find they can couple to various structures, including hidden-charm baryonium states, charmonium states plus two pions, and hidden-charm tetraquark states plus one pion, etc. At the beginning we do not know which structure these currents couple to, but after sum rule analyses we can obtain some information. We find some of them can couple to hidden-charm baryonium states, using which we evaluate the masses of the lowest-lying hidden-charm baryonium states with quantum numbers $J^P=2^-/3^-/0^+/1^+/2^+$ to be around 5.0 GeV. Our results suggest:
\begin{enumerate}

\item The hidden-charm baryonium currents can probably couple to both charmonium states plus two pions [$\bar cc + \pi\pi$] and hidden-charm baryonium states [$\bar cc\bar qq \bar qq$], and may couple to hidden-charm tetraquark states plus one pion [$\bar cc\bar qq$+$\pi$]. Actually, the mass signal around 3.5 GeV in the left panel of Fig.~\ref{fig:typeB} is quite clear, probably related to the $\eta_c \pi \pi$ and $J/\psi \pi \pi$ thresholds.

\item All the currents of Type D with the positive parity and all the currents of Type F with the negative parity lead to reasonable sum rules. The masses of the lowest-lying hidden-charm baryonium states are evaluated to be: 4.83~GeV~($J^P=2^-$), 5.04~GeV~($J^P=3^-$), 5.00~GeV~($J^P = 0^+$), 4.89~GeV~($J^P = 1^+$), 5.15~GeV~($J^P = 2^+$), and 5.68~GeV~($J^P = 3^+$). According to their spin-parity quantum numbers and Eq.~(\ref{eq:cr}), the hidden-charm baryonium states of $J^P=2^-/3^-$ may be observed in the $D$-wave $\eta_c \pi \pi$ and $J/\psi \pi \pi$ channels, and those of $J^P=0^+/1^+/2^+$ may be observed in the $S$-wave $J/\psi \rho$ and $J/\psi \omega$ channels. More information on their isospin, etc. will be discussed in our future studies.

\item All the currents of $J^P=0^-$ and $1^-$ seem not to couple to hidden-charm baryonium states, possibly because these currents can be significantly affected by the $S$-wave $\eta_c\pi\pi$ and $J/\psi\pi\pi$ thresholds. However, we do not want to conclude that there do not exist hidden-charm baryonium states of $J^P=0^-$ and $1^-$, but note that these states might not be easily observed if their coupling to $\eta_c\pi\pi$ and $J/\psi\pi\pi$ channels can not be well constrained.

\end{enumerate}

To end this paper, we note that in the present study we have constructed many interpolating currents, some of which have the same quantum numbers and can mix with each other. Moreover, the physical states are not one-to-one related to these currents, and are probably a mixture of various components coupled by relevant currents. Basing on this partial coupling, in this paper we have studied the hidden-charm baryonium states using the method of QCD sum rule, but to study them in a more exact way, one needs to calculate both the diagonal elements (the two-point correlation function using the same current) and the off-diagonal elements (the two-point correlation function using two different currents). However, we have only done the former calculations in the present study, because the latter still seem difficult in the current stage and wait to be solved.

Based on the results of this letter, we suggest to search for the hidden-charm baryonium states, especially the one of $J=3^-$, in the $D$-wave $J/\psi \pi \pi$ and $P$-wave $J/\psi \rho$ and $J/\psi \omega$ channels in the energy region around 5.0 GeV, and hope they can be discovered with the running of LHC at 13 TeV and forthcoming BelleII in the near future. We would also like to see other theoretical studies which can give more accurate predictions on their masses, productions and decay properties, etc.

\section*{Acknowledgments}
\begin{acknowledgement}
This project is supported by the Natural Sciences and Engineering Research Council of
Canada (NSERC) and the National Natural Science Foundation of China under Grants
No. 11475015, No. 11375024, No. 11222547, No. 11175073, and No. 11575008; the Ministry of
Education of China (SRFDP under Grant No. 20120211110002 and the Fundamental Research
Funds for the Central Universities); the National Program for Support of Top-notch Youth Professionals.
\end{acknowledgement}


\end{document}